\documentclass[useAMS,usenatbib,usegraphicx]{mn2e}
\usepackage{epsfig}
  
\begin{document}

\title[Simulated Colour-Magnitude Diagrams]{Interpreting the Colour-Magnitude Diagrams 
of Open Star Clusters through Numerical Simulations}
\author[J. S. Kalirai and M. Tosi]{Jasonjot Singh Kalirai$^{1}$\thanks{E-mail:
jkalirai@physics.ubc.ca; tosi@bo.astro.it} and Monica 
Tosi$^{2}$\footnotemark[1] \\
$^{1}$Department of Physics \& Astronomy, 6224 Agricultural Road, University 
of British Columbia, Vancouver BC V6T1Z1\\
$^{2}$INAF - Osservatorio Astronomico di Bologna, Via Ranzani 1,
I-40127 Bologna, Italy}

\date{Accepted 2004 March 08. Received 2004 January 13; in original form 2004 
January 13}

\pagerange{\pageref{firstpage}--\pageref{lastpage}} \pubyear{2004}

\maketitle

\label{firstpage}

\begin{abstract}
We present detailed comparisons between high quality  
observational colour-magnitude diagrams (CMDs) of open star clusters 
and synthetic CMDs based on MonteCarlo numerical simulations.  The 
comparisons account for all of the main parameters which determine 
the shape of the CMD for a stellar population.  For the four clusters 
studied, NGC 6819, NGC 2099 (M37), NGC 2168 (M35) and 
NGC 2323 (M50), we derive reddening, distance, age, binary fraction, star 
formation rate and indicative metallicity by comparing the locations and 
density of points in the observed CMDs to the simulated CMDs.  
We estimate the uncertainties related to stellar evolution theories by adopting
various sets of stellar models for all of the synthetic CMDs and discuss which 
stellar models provide the theoretical CMDs that best reproduce the 
observations.
\end{abstract}

\begin{keywords} colour-magnitude diagrams -- methods: n-body simulations -- 
open clusters and associations: general -- open cluster and associations: 
individual: NGC 2099, NGC 2168, NGC 2323, NGC 6819
\end{keywords}

\section{Introduction}\label{intro}

	Theoretical isochrones are commonly fit to the major observational  
sequences of star clusters in order to both better understand the 
underlying physics of stellar evolution and to determine properties 
of the clusters, i.e., the age.  If the metallicity, reddening and distance 
to the cluster are well constrained from independent techniques, the 
comparisons typically involve matching the morphology of the turn-off and 
location of the red giant stars to predictions.  Recently, the newer method 
of using synthetic colour-magnitude diagrams to compare with observational 
data has proven to be much more informative and rewarding 
\citep{aparicio,tosi,skillman}.  These MonteCarlo simulations allow modelling of 
several additional parameters which dictate the distribution of points 
in the CMD, such as stochastic star formation (SF) processes, binary fraction, 
photometric spread, main-sequence thickness, data incompleteness and small 
number statistics.  Consequently, the results not only provide a measure of 
the properties of the cluster, but can also constrain the star formation 
history (SFH) and the initial mass function (IMF).  
Furthermore, by comparing the simulations based on several different sets of 
evolutionary tracks, we can constrain which models use the best prescription 
of parameters (such as treatment of overshooting, mixing length, etc...).

	Confronting the simulations with observations requires a large data 
set with accurate photometry.  For this, we use the deep $BV$ photometry 
presented in the CFHT Open Star Cluster Survey (\cite{kalirai1}, hereafter 
JSKI).  JSKI observed 19 open star clusters in our Galaxy and have 
yet published results on the four richest clusters, NGC 6819 (\cite{kalirai2}, 
hereafter JSKII), NGC 2099 (\cite{kalirai3}, hereafter JSKIII), and 
NGC 2168 and NGC 2323 (\cite{kalirai4}, hereafter JSKIV).  These data were 
reduced and calibrated in a homogenous manner as described 
in JSKI.  The resulting CMDs exhibit very tight main sequences showing 
several `kinks' and slope changes which are predicted by theory.  More 
importantly, the combination of very short and deep exposures, and the 
large aerial coverage of the detector (42$' \times$ 28$'$) has allowed the 
measurement of stars from the brightest asymptotic giant branch (AGB) and red 
giant branch (RGB) phases down to very low-mass main-sequence phases ($\sim$0.2 
M$_\odot$).  This allows our comparisons to yield evolutionary information over 
a wide mass range.

	The reduced data set in the CFHT Open Star Cluster Survey has been 
requested by, and made available to, several investigators for additional 
science rewards outside our goals (e.g., astrometric studies, proper motions,
variable stars, radial velocities, brown dwarfs, blue stragglers and Galactic 
disk star distributions).  The present study complements these efforts and
analyses the four published clusters in a way that allows us to include them 
in a large homogeneous sample of open clusters aimed at studying the formation 
and evolution of the Galactic disk (Bragaglia 2003, and references therein). 
Galactic open clusters are indeed particularly 
well suited to this purpose, since they span a range of ages from a few million 
to several billion years and can be observed in various regions of the Galactic
disk characterised by different star formation histories. They can be used to
study both the present day disk structure and its temporal evolution
(Janes \& Phelps 1994, Friel 1995, Tosi 2000, Bellazzini et al. 2003). Old
open clusters offer a unique opportunity to trace the whole kinematical and
chemical history of our disk, if collected in populous and representative  
samples and accurately and homogeneously analysed (see e.g., Twarog, Ashman, 
\& Anthony-Twarog 1997; Carraro, Ng, \& Portinari 1998).

Here we apply the synthetic CMD method to NGC 6819, NGC 2099, NGC 2168 and NGC
2323 to derive their age, reddening, distance modulus and (approximate)
metallicity homogeneously to the Bragaglia (2003) cluster sample. The
method also allows us to determine other features of these clusters, such as
the existence (or lack thereof) of a significant fraction of unresolved binary
systems, the original total mass of formed stars and the possible evaporation
of some of the lower mass stars.

	The organisation of the paper is as follows, \S \ref{observations} 
briefly summarises the data and the reduction procedures.  Further 
details are given in JSKI.  In \S \ref{firstresults} we present a 
summary of our main results which relate to this work from the published 
papers in the CFHT Open Star Cluster Survey.  \S \ref{synthetic} sets up 
the numerical simulations and presents details on how the synthetic CMDs 
were created.  Next, we compare the CMDs and the corresponding luminosity
functions from the observations with the simulations on a 
cluster-by-cluster basis (\S \ref{vs}).   Finally, we discuss 
the results in \S \ref{discussion} and conclude the study in \S 
\ref{conclusions}.

\section{The Observational Data}\label{observations}

	An absolute requirement in using different data sets to 
make detailed comparisons between observations and models is that 
of homogeneity.  To truly understand the variations in the different models
parameters,  we must minimize systematic errors.  Therefore, we desire a rich 
data set of open clusters, all reduced using the same approach,  
calibrated consistently, and with well determined incompleteness factors.
The richest clusters from the CFHT Open Star Cluster Survey, NGC 6819, NGC 2099 (M37), 
NGC 2168 (M35) and NGC 2323 (M50) perfectly meet these requirements. The data were all 
imaged during an excellent three night 
observing run with the Canada-France-Hawaii Telescope (CFHT) on 1999 
October, using the CFH12K camera.  The optical detector is a high 
resolution (1$\times$10$^8$ pixels) CCD mosaic camera which projects to 
an aerial coverage of 42$'$$\times$28$'$ on the sky.

\begin{table*}
\caption{Observational Data for NGC 6819, NGC 2099 (M37), NGC 2168 (M35) and NGC 
2323 (M50)}
\begin{tabular}{cccccc}
\hline
Cluster & Filter & Exposure Time (s) & No. of Images & Seeing ($''$) & Airmass \\
\hline

NGC 6819 ...... \\

& $V$ & 300 & 9 & 0.7 & 1.3 \\

& $V$ & 50 & 1 & 0.7 & 1.16 \\

& $V$ & 10 & 1 & 0.68 & 1.15 \\

& $V$ & 1 & 1$^{*}$ & 0.78 & 1.27  \\

& $B$ & 300 & 9 & 0.9 & 1.40-1.76 \\

& $B$ & 50 & 1 & 0.82 & 1.38 \\

& $B$ & 10 & 1 & 0.84 & 1.37 \\

& $B$ & 1 & 1$^{*}$ & 1.1 & 1.25   \\

\hline

NGC 2099 ...... \\

& $V$ & 300 & 3 & 0.85 & 1.03 \\

& $V$ & 50 & 1 & 0.97 & 1.04 \\

& $V$ & 10 & 1 & 0.99 & 1.04 \\

& $V$ & 0.5 & 1$^{*}$ & 1.1 & 1.19  \\

& $B$ & 300 & 3 & 0.79 & 1.03 \\

& $B$ & 50 & 1 & 0.85 & 1.03 \\

& $B$ & 10 & 1 & 0.85 & 1.03 \\

& $B$ & 0.5 & 1$^{*}$ & 1.0 & 1.19   \\

\hline

NGC 2168 ...... \\

& $V$ & 300 & 1 & 1.35 & 1.6 \\

& $V$ & 50 & 1 & 1.35 & 1.6 \\

& $V$ & 10 & 1 & 1.35 & 1.6 \\

& $V$ & 1 & 1$^{*}$ & 1.2 & 1.35  \\

& $V$ & 0.5 & 1$^{*}$ & 1.2 & 1.35  \\

& $B$ & 300 & 1 & 1.2 & 1.6 \\

& $B$ & 50 & 1 & 1.2 & 1.6 \\

& $B$ & 10 & 1 & 1.2 & 1.6 \\

& $B$ & 1 & 1$^{*}$ & 1.1 & 1.35   \\

& $B$ & 0.5 & 1$^{*}$ & 1.1 & 1.35  \\

\hline

NGC 2323 ...... \\

& $V$ & 300 & 1 & 0.85 & 1.15 \\

& $V$ & 50 & 1 & 0.85 & 1.15 \\

& $V$ & 10 & 1 & 0.85 & 1.15 \\

& $V$ & 1 & 1$^{*}$ & 1.0 & 1.2  \\

& $V$ & 0.5 & 1$^{*}$ & 1.0 & 1.2  \\

& $B$ & 300 & 1 & 0.95 & 1.15 \\

& $B$ & 50 & 1 & 0.95 & 1.15 \\

& $B$ & 10 & 1 & 0.95 & 1.15 \\

& $B$ & 1 & 1$^{*}$ & 1.1 & 1.2  \\

& $B$ & 0.5 & 1$^{*}$ & 1.1 & 1.2  \\
\hline

\end{tabular}

1. $^{*}$ These very short exposures were obtained at a later date. \\

\end{table*}
	 
Most clusters were observed with exposure times sufficient to
reach the magnitude of the coolest white dwarfs and of the faintest M-type stars 
on the main sequence.    We also complemented the deep 
photometry with shorter exposures to cover the brighter turn-off 
and evolved stars.  The individual images were registered and co-added 
(where necessary) using the FITS Large Images Processing Software (FLIPS) (J.-C. 
Cuillandre, 2000, Private Communication).  Details of this procedure, as well 
as the pre-processing of the data are provided in \S 3 of JSKI.  

	The actual photometry of the sources in the images was measured using 
Point-Spread-Function Extractor (PSFex), a highly automated program which 
will be integrated in the code of a future SExtractor version (Bertin \& Arnouts 
1995; E. Bertin 2000, private communication).  We used a separate PSF on each 
CCD of the mosaic.  Included in the PSF are polynomial basis functions which 
can map the variations of the PSF across the CCD.  PSFex is discussed further 
in \S 4 of JSKI.
	
Data for each of the clusters were taken in the $B$ and $V$ filters, 
and calibrated in the Johnson photometric system using multiple images of the 
SA-92 and SA-95 standard star fields \citep{landolt}.   
The final averaged uncertainties in the 
zero-point terms were measured to be $\sim$0.018 magnitudes, uncertainties 
in the airmass coefficients were $\sim$0.015 magnitudes in $V$ and $\sim$0.007 
magnitudes in $B$.  Colour coefficients, as well as airmass terms, were similar 
to nominal CFHT expectations.  Details of the calibration of instrumental 
magnitudes to real magnitudes are provided in \S 5 and Tables 2 and 3 of JSKI.

	The resulting, calibrated CMDs for the open clusters exhibit 
unprecedented long, clear main sequences ranging from the brightest cluster 
members to the limiting magnitude in each case (23 $\leq$ $V \leq$ 25).  
A summary of the observational log for each cluster (exposure time, 
seeing, airmass, etc...) is presented in Table 1.

\section{First Results: Isochrone Fitting }\label{firstresults}

Figure 1 presents the final calibrated CMDs for each of the four 
open clusters, with the best fit theoretical isochrones as determined 
in our previous published analysis of these clusters (see JSKII, JSKIII, JSKIV
for details).  
These rich clusters are clearly delineated from the 
background Galactic disk stars.  The main sequences are very tight 
and show several features which are predicted by stellar evolutionary 
theory (see \S 5).   For this, we 
used a new grid of model isochrones calculated by the group at the Rome 
Observatory according to the prescription described in \cite{ventura}.  
These models adopt convective core-overshooting by means 
of an exponential decay of the turbulent velocity out of the formal 
convective borders fixed by Schwarzschild's criterion; a free 
parameter which gives the e-folding distance of 
the exponential decay is set to $\zeta$ = 0.03 \citep{ventura}.  
The convective flux has been evaluated
according to the Full Spectrum of Turbulence (FST) theory
prescriptions \citep{canuto}. The theoretical isochrones were 
transformed into the observational plane by making use of 
the Bessel, Castelli \& Pletz (1998) conversions.  The lower main sequence 
(M $\la$ 0.7 M$_\odot$) was calculated 
by adopting NextGen atmosphere models \citep{hauschildt}. 
  For M $\la$ 0.47 M$_\odot$ (or 
T$_{\rm eff}$ $\la$ 3500 K) the transformations 
of \cite{hauschildt} in $B{\rm-}V$ are not very reliable and so 
the faint end of the isochrones terminate at this mass.  These models will 
be referred to as FST from here on.

 
\begin{figure*}
\centering
\epsfig{file=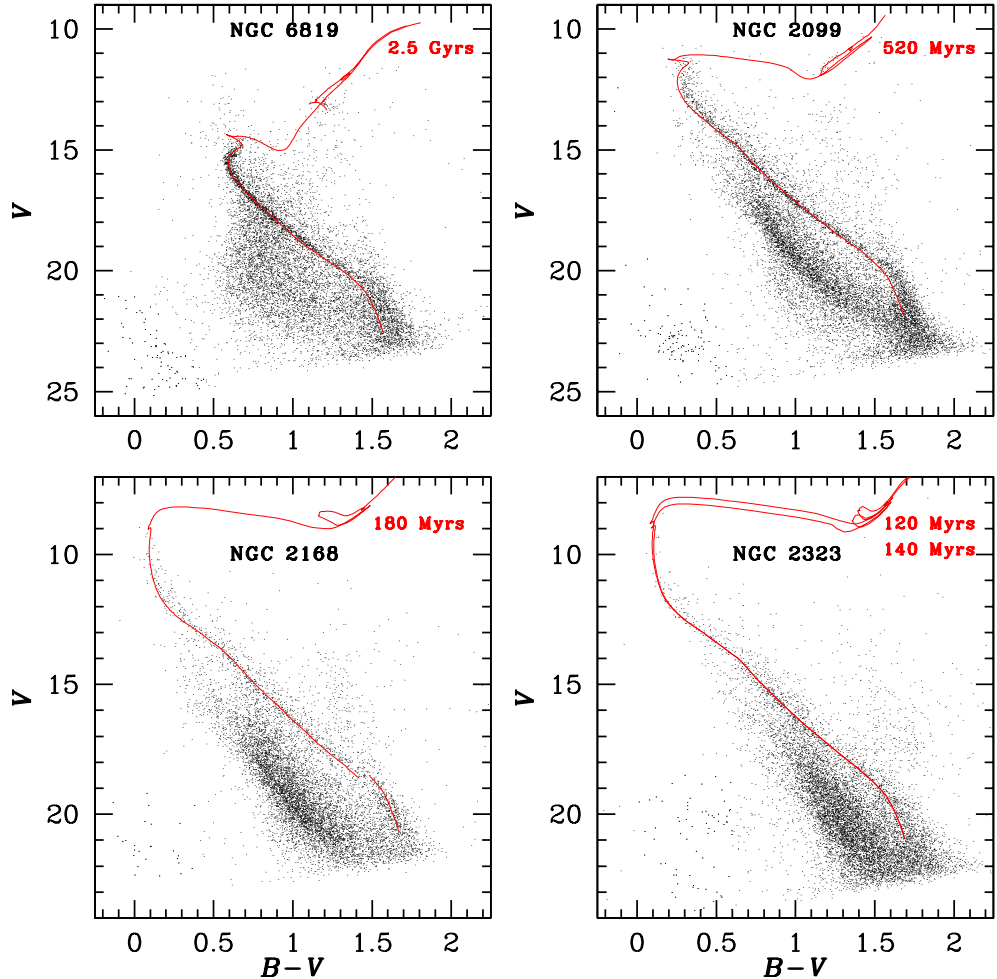,width=0.75\textwidth}
\caption{Isochrone fits for each cluster are shown with corresponding ages.  All 
models are Solar metallicity with the exception of NGC 2168, for which $Z$ = 0.012. 
For the lower main sequence, we were not able to compute non-grey atmospheres for 
this metallicity.  We therefore overplot the lower main sequence from a Solar 
metallicity model (bottom-left).}
\label{CMDTheory}
\end{figure*}


\subsection{Published Results -- NGC 6819}\label{preresults6819}

The tight, very rich, main sequence and turn-off consist of over 2900 
cluster stars to our limiting magnitude. Main-sequence
fitting of the un-evolved cluster stars with the Hyades star
cluster yields a distance modulus of ($m{\rm-}M$)$_{V}$ = 12.30
$\pm$ 0.12 ($d$ = 2500 pc), for a reddening of $E$($B{\rm-}V$) = 0.10. 
These values are consistent with a theoretical isochrone of 
age 2.5 Gyrs.  Detailed star
counts in concentric annuli out to large angular radii set constraints on 
the cluster radius, $R$ = 9$\farcm$5 $\pm$1$\farcm$0, and clearly indicates 
mass segregation in the cluster.  The global cluster mass function is 
found to be quite flat, $x$ = {$\rm-$}0.15 ($x$ = 1.35, Salpeter).  A large 
population of white dwarf stars is found in the cluster and is currently 
being investigated spectroscopically.

\subsection{Published Results -- NGC 2099 (M37)}\label{preresults2099}

The cluster CMD shows 
extremely well populated and very tightly constrained main sequence,
turn-off, and red giant populations.  The photometry for this cluster is 
faint enough ($V \sim$ 24.5) to detect the end of the white dwarf cooling 
sequence.  Therefore, the white dwarfs are used as chronometers to age 
the cluster providing an independent age (566 $\pm \ ^{154}_{176}$ Myrs) 
from the main-sequence turn-off plus red giant clump fit (520 Myrs).  We 
also derive the reddening (E($B{\rm-}V$) = 0.21 $\pm$ 0.03) and distance 
(($m{\rm-}M$)$_{V}$ = 11.55 $\pm$ 0.13) to NGC 2099 by matching 
main-sequence features in the cluster to a fiducial main-sequence for 
the Hyades \citep{deBruijne}, after correcting for small metallicity 
differences.  The cluster luminosity and mass functions indicate some 
evidence for mass segregation within the boundary of the cluster.

\subsection{Published Results -- NGC 2168 (M35)}\label{preresults2168}

NGC 2168 is a well studied nearby (d = 912 $\pm \ ^{70}_{65}$ pc) young 
cluster (180 Myrs) showing a well defined main sequence. 
 The reddening of the cluster, 
$E$($B{\rm-}V$) = 0.20, and the metallicity, $Z$ = 0.012, were adopted from 
the literature \citep{sarrazine, barrado}.  The cluster is found to contain 
$\sim$1000 stars above our limiting magnitude with a global mass function 
slope very similar to a Salpeter value ($x$ = 1.29).  There is mild evidence for 
mass segregation in the cluster.  White dwarf stars are found and can potentially 
set very important constraints on the high-mass end of the white dwarf initial-final 
mass relationship considering they must have evolved from quite massive progenitors.

\subsection{Published Results -- NGC 2323 (M50)}\label{preresults2323}

NGC 2323 is one of the youngest clusters in the CFHT Open Cluster Survey 
(130 Myrs), and shows 
a main sequence extending over 14 magnitudes in the $V$, $B{\rm-}V$ 
plane.  The cluster has had very few photographic/photoelectric studies, 
and prior to our efforts, no previously known CCD analysis.  This is surprising 
considering the rich stellar population ($\sim$2100 stars above our magnitude 
limit) and the relatively low distance modulus ($(m{\rm-}M$)$_\circ$ = 10.00 
$\pm$ 0.17).  The cluster shows clear evidence of mass segregation, an important 
result considering the dynamical age is only 1.3$\times$ the cluster age.

\section{Synthetic Colour-Magnitude Diagrams} \label{synthetic}

Age, reddening and distances for the four clusters have been re-derived applying 
the synthetic CMD method (Tosi et al. 1991) to the empirical CMDs described 
above. The best values of the parameters are found by
selecting the values that provide synthetic CMDs with morphology, number of stars 
in the various evolutionary phases and luminosity functions in better 
agreement with the empirical ones. The method has already been succesfully 
applied to several clusters (Bragaglia 2003 and references therein).

The synthetic CMDs are constructed via MonteCarlo extractions
of (mass, age) pairs, according to an assumed IMF, SF law, and time
interval of the SF activity. Each extracted synthetic star is placed in
the CMD by suitable interpolations on the adopted stellar evolution tracks
and adopting the Bessel et al. (1998) tables for photometric conversion in 
the Johnson-Cousins photometric system. The absolute magnitude is converted 
to a {\it provisional} apparent magnitude by applying (arbitrary) reddening and 
distance modulus. The synthetic stars extracted for any magnitude and photometric 
band are assigned the photometric error derived for the actual stars of the 
same apparent magnitude. Then, they are randomly retained or rejected on the 
basis of the incompleteness factors of the actual data, derived from extensive
artificial star tests. 

Once the number of objects populating the whole synthetic CMD (or portions
of it) equals that of the observed one, the procedure is stopped, yielding the
quantitative level of the SF rate consistent with the observational data,
for the prescribed IMF and SF law. 
To evaluate the goodness of the model predictions, we compare them with:
the observational luminosity functions, the overall morphology of the
CMD, the stellar magnitude and colour distributions, the number of objects in particular
phases (e.g., on the red giant branch, in the clump, on the blue loops, etc.).
A model can be considered satisfactory only if it reproduces all of the features
of the empirical CMDs and luminosity functions. Given the uncertainties affecting both the
photometry and the theoretical parameters (stellar evolution tracks included),
the method cannot provide strictly unique results; however, it 
allows us to significantly reduce the range of acceptable parameters.

In this way we derive the age, reddening, distance
of the cluster, and indicate the metallicity of the stellar evolution models in
better agreement with the data.  In principle, the latter should be indicative 
of the cluster metallicity, but this depends significantly on some of the stellar 
model assumptions, such as opacities.

To estimate if, and how many, unresolved binary systems could be present in the
cluster, the synthetic CMDs have been computed both assuming that all the 
cluster stars are single objects and that an (arbitrary) fraction of them are 
members of binary systems with random mass ratio. Again, the comparison of the
resulting main-sequence morphology and thickness with the observed ones allows us to
derive information on the most likely fraction of binaries.

We also infer the astration mass of the cluster (i.e., the total mass that went 
in all of the stars formed in the cluster), according to the adopted IMF.
The method here also allows us to derive the best IMF as long as the observed mass 
range of main-sequence stars is sufficiently large and the main sequence is 
sufficiently tight. However, the high background contamination affecting the systems 
examined here and the lack of sufficiently reliable external blank fields, make 
the derivation too unsafe to discuss it here.  Other factors (see below), such 
as the actual size of the photometric errors also prevents us from 
comparing the IMF in any significant detail.

\section{Observations vs Theory - Colour-Magnitude Diagrams and Luminosity
Functions} \label{vs}

To test the effect of different input physics on the derived  
 age, reddening and distance modulus, we have run the simulations with 
three different sets of stellar evolutionary tracks.  The adopted sets were 
chosen because they assume different prescriptions for the treatment of 
convection and range from no overshooting to rather high overshooting from 
convective regions. Despite these differences, they are all able to well
reproduce the observed CMDs of both star clusters and nearby galaxies.  
They are thus suitable to evaluate the intrinsic uncertainties still 
related to stellar evolution models.
We use the FST tracks of \cite{ventura} with high ($\zeta$ = 0.03) and moderate
($\zeta$ = 0.02) overshooting;   the BBC94 tracks of the Padova group 
\citep{bressan,fagotto}, with overshooting; and the FRANEC tracks 
\citep{dominguez} with no overshooting.  
For those cases where the metallicities are not well constrained, we allow 
different values in our comparisons.  Incompleteness factors and photometric 
errors are taken from the original papers in the CFHT Open Star Cluster Survey 
series (JSKII, JSKIII and JSKIV) and folded into the numerical simulations.

All models assume that the star formation activity has lasted 5 Myr (i.e., 
approximately an instantaneous burst) and that the stars were formed following
a single slope ($x$ = 1.35) Salpeter's IMF over the whole mass range covered
by the adopted tracks (0.6 -- 100 $M_{\odot}$). 

Figures 2, 4, 6 and 8 show for each cluster the synthetic models in better 
agreement with the data for each adopted set of stellar tracks, as well as the 
cluster observational CMD on the same scale. In all of these figures the CMDs in 
the top row are the synthetic ones, those in the second row are the same
synthetic ones but overlapped with the stars of the equal area fiducial blank
field, and the bottom CMD is the empirical one. Figures 3, 5, 7 and 9 compare
the empirical luminosity functions with those predicted by the best synthetic 
models.  Ultimately, the best models have been selected by visually comparing 
features between the synthetic models and the observations, such as the location 
and density of points on the CMD.  However, we have also attempted to confirm 
our results by using both the two-dimensional Kolmogorov-Smirnov test 
on the CMD and $\chi^2$ tests on the luminosity and colour 
functions.  In the cases of well constrained field contamination, such as behind 
NGC 6819 and NGC 2099, the tests confirm our choices of best synthetic models.  
For the other two clusters, the tests do not help discriminate between best 
choice of models and we strictly use conclusions based on the visual location 
of points.  Also, these statistical tests are particularly not well suited for 
these comparisons because of the insignificant weight given to the small 
number of stars in key evolutionary stages which must be fit by the synthetic 
models (such as the red giant clump).

\subsection{NGC 6819} \label{ngc6819}


\begin{figure*}
\centering
\epsfig{file=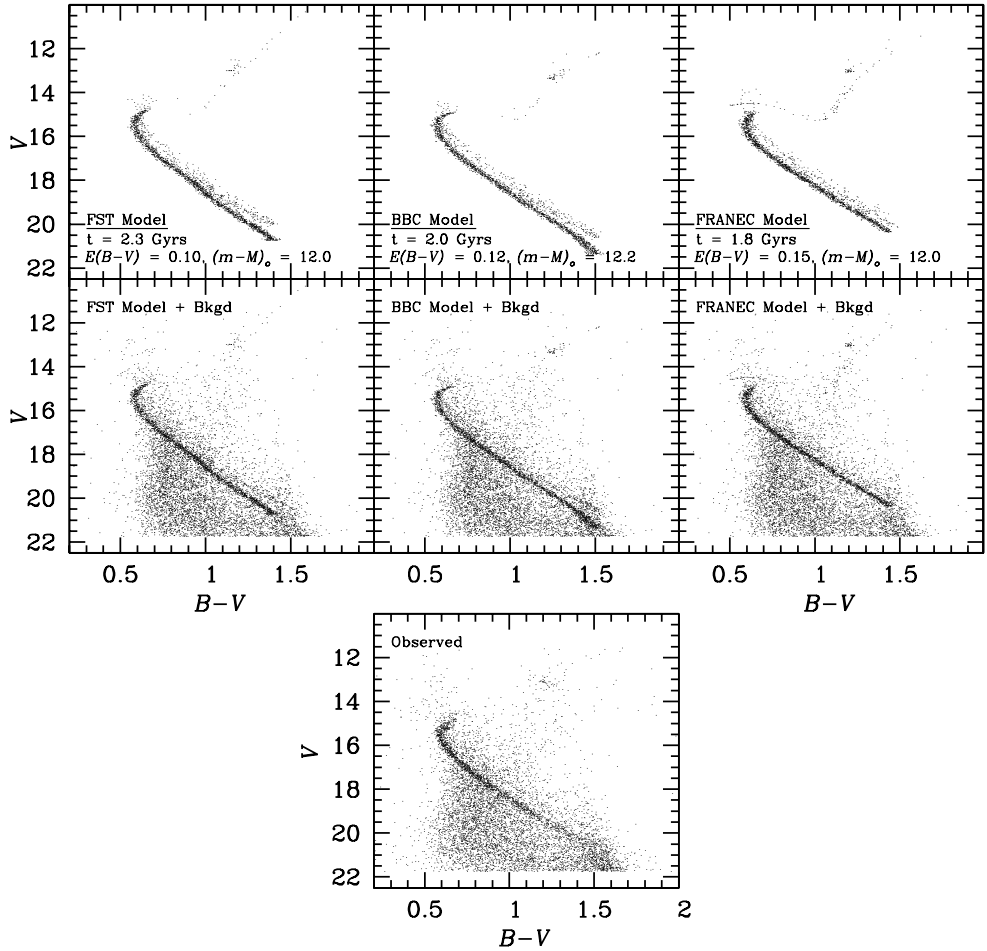,width=0.75\textwidth}
\caption{Synthetic CMDs with $Z$ = 0.02 and a 20\% binary 
fraction (random mass ratio) are shown for NGC 6819.  The BBC94 tracks are found 
to provide the best agreement with the observations (bottom panel).  The data has 
been truncated at the faintest magnitudes (see Figure 1 for full CMD).  See \S 
\ref{ngc6819} for more information.}
\label{N6819cmd}
\end{figure*}


NGC 6819 is the richest cluster in this work. The cluster field contains 10510
{\it bona fide} stars (i.e., with {\it stellarity} parameter $\geq$ 0.5) 
while the blank field of same area contains 8504 {\it bona fide} stars. 
The synthetic CMDs are therefore computed with 2000 objects (a fraction of
which are assumed to be members of binary systems).

All of the tracks used to simulate the NGC 6819 CMD are Solar metallicity models as 
the cluster metallicity has been constrained to this value 
through the high resolution analysis of \cite{bragaglia}.  
Figure 2 shows that the 
three sets of models provide slightly different ages (and hence, reddening and 
distances), which strictly depend on the assumed amount of overshooting: 
the higher the overshooting, the older the age. This is of course due to higher 
overshooting models having higher stellar luminosities, thus allowing
lower mass stars to  reach the desired luminosity.  
The variations in the ages seen in the models 
also stress the importance of deriving uncertainties on cluster parameters 
by using different sets of tracks.  


\begin{figure*}
\centering
\epsfig{file=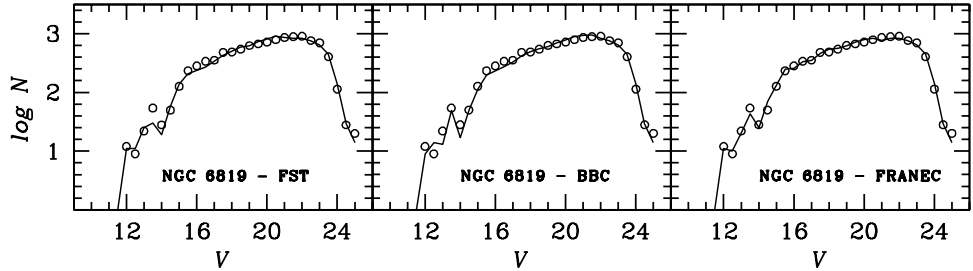,width=0.75\textwidth}
\caption{The empirical NGC 6819 luminosity function (open circles) 
is compared with the predicted luminosity function from the numerical simulations 
(solid line).  See \S \ref{ngc6819} for a discussion of these results.}
\label{N6819lf}
\end{figure*}


The FST models with $\zeta$ = 0.03 (left) lead to parameters virtually 
identical to those  found in JSKII.  
A better estimate of the age is found to be 2.30 $\pm$ 0.15 Gyrs, 
distance modulus $(m {\rm-}M)_\circ$ = 12.0, and reddening $E(B {\rm-}V)$ = 0.10.  The 
FRANEC models (right) suggest younger ages, due to their lack of overshooting, 
1.6-1.8 Gyrs, and 
therefore need a slightly higher reddening ($E(B {\rm-}V)$ = 0.15) to reproduce 
the colours.  The BBC94 models (middle), with intermediate overshooting, favor an 
intermediate age of 2.0 Gyrs, with $E(B {\rm-}V)$ = 0.12 and $(m {\rm-}M)_\circ$ = 12.2.

None of these models are found to perfectly reproduce all of the observed features in 
the NGC 6819 CMD.  For example, the FST models predict a main-sequence 
slope steeper than the observed one and the two diverge at the faint end. These models 
also tend to place the red giant clump excessively towards the red, probably due 
to the fact that the models don't include mass loss.  The FRANEC models better reproduce 
the main-sequence distribution but present a turn-off with a hook that is too 
pronounced, a likely consequence of the young age.  Given the high field 
contamination, however, this feature is not particularly evident in the 
combined CMDs of the second row.  A 
smaller cluster core CMD shows the hook prominently.  The FRANEC models also 
overpopulate the lower red giant branch.  This is visible even despite the 
fact that this region of the CMD is plenty of contaminating objects.
The BBC94 models provide the best fit to the data CMD, despite a main-sequence 
slope which is slightly too steep.  Most of these discrepancies between the 
observed features and the synthetic CMDs result from problems in the stellar
models.  As pointed out by Andreuzzi et al. (2004) studying the open cluster
NGC 6939 with the same method, assuming a reddening dependence on the stellar
colour as adopted by Twarog et al. (1997) makes the synthetic main sequences even 
steeper and, hence, more inconsistent with the data. 

Another evident result that we find for all of the clusters studied here, is that the 
observed main-sequence spread cannot be reproduced if one assumes photometric 
errors as small as those listed in our photometry catalogue.  To test the actual 
size of the photometric errors we revert back to the artificial star tests 
presented in JSKII.  A plot of the $V_{\rm in}$ {\rm-} $V_{\rm out}$ distribution 
tells us that, in fact, the errors quoted agree with the artificial star tests.  
The synthetic CMDs however, require that either the errors are larger, at least 
0.01 magnitude even in the brighter bins, or there is some physical cause for the 
observed spread (i.e., differential reddening or metallicity spread).  A spread due 
to binaries is modelled and therefore ruled out as a cause for this discrepency. 
The apparent spread could be caused by small offsets in the calibration 
of the different CCDs of the mosaic, but internal metallicity spread 
in the cluster or small differential reddening cannot be excluded.  
We are currently investigating this possibility in 
another project (R. S. French et al. 2003, private communication).  For now, we have 
increased the errors to 0.01 magnitude for stars brighter than $V$ = 19 and doubled 
them for fainter bins ($\sigma$ = 0.01 at $V$ = 21, $\sigma$ = 0.04 at $V$ = 23).  
Empirically, this is the minimum size required to obtain a main-sequence spread 
comparable with the observed one.

All of the shown cases in Figure 2 assume that 20\% of the stars are in binary 
systems with random mass ratio.  We also tested simulations with 30\% 
binaries and found minor differences.  Given the high field contamination, it 
is difficult to understand which fraction and assumption of the mass ratio is 
the best case. It could entirely be true that the fraction is actually smaller 
if the mass ratio is more skewed towards equal masses.  Models without
binaries are not shown as they amplify the problem of the main-sequence width 
(the predicted one is too tight) and do not correctly reproduce the stellar 
distribution on the red side of the main sequence.

Figure 3 shows the empirical luminosity function of the 10510 cluster field stars (empty circles) 
compared to the luminosity functions resulting from the sum of the 2000 synthetic stars of 
Figure 2 with the 8504 stars of the blank field.  The agreement is excellent,
except for the slight underestimate of the clump stars of the FST models. 

A clear result of our study is that all of the synthetic CMDs for NGC 6819 show 
overpopulated
faint main sequences ($V >$ 20), both for single and binary stars.  This is a 
clear signature of low mass stellar evaporation (or at least segregation 
outwards of our cluster boundary, 9$\farcm$5).  As we showed in JSKII, the 
mass function of this cluster steadily marches to flatter values, and becomes 
inverted in the outermost annuli.  Therefore, the observations provide the 
answer to this overabundance.  However the effect is not seen in the luminosity 
function comparisons, which are in excellent agreement with the 
observed one.  This suggests either that the faint magnitude luminosity function 
is dominated by field contamination or that the apparent over-concentration of the 
synthetic main sequences is actually an artifact of the residual underestimate of 
the photometric errors.

Assuming a single slope Salpeter IMF, the mass of gas that formed the stars of
the surveyed area of NGC 6819 between 0.6 and 100 $M_{\odot}$ is 
$\simeq 4.2 \times 10^3 M_{\odot}$ with the BBC94 models, $4.0 \times 
10^3 M_{\odot}$ and $4.9 \times 10^3 M_{\odot}$ with the FST and the FRANEC
models, respectively. Hence, different stellar models do not provide the same 
answer for a global parameter such as the star formation either. 
The differences are again related to the adopted input physics and to the
amount of assumed overshooting in the stellar models. Less overshooting 
implies higher masses for the observed living stars, i.e. shorter lifetimes,
which, convolved with the negative slope of the IMF, leads to a higher total 
astrated mass. 

\subsection{NGC 2099} \label{ngc2099}


\begin{figure*}
\centering
\epsfig{file=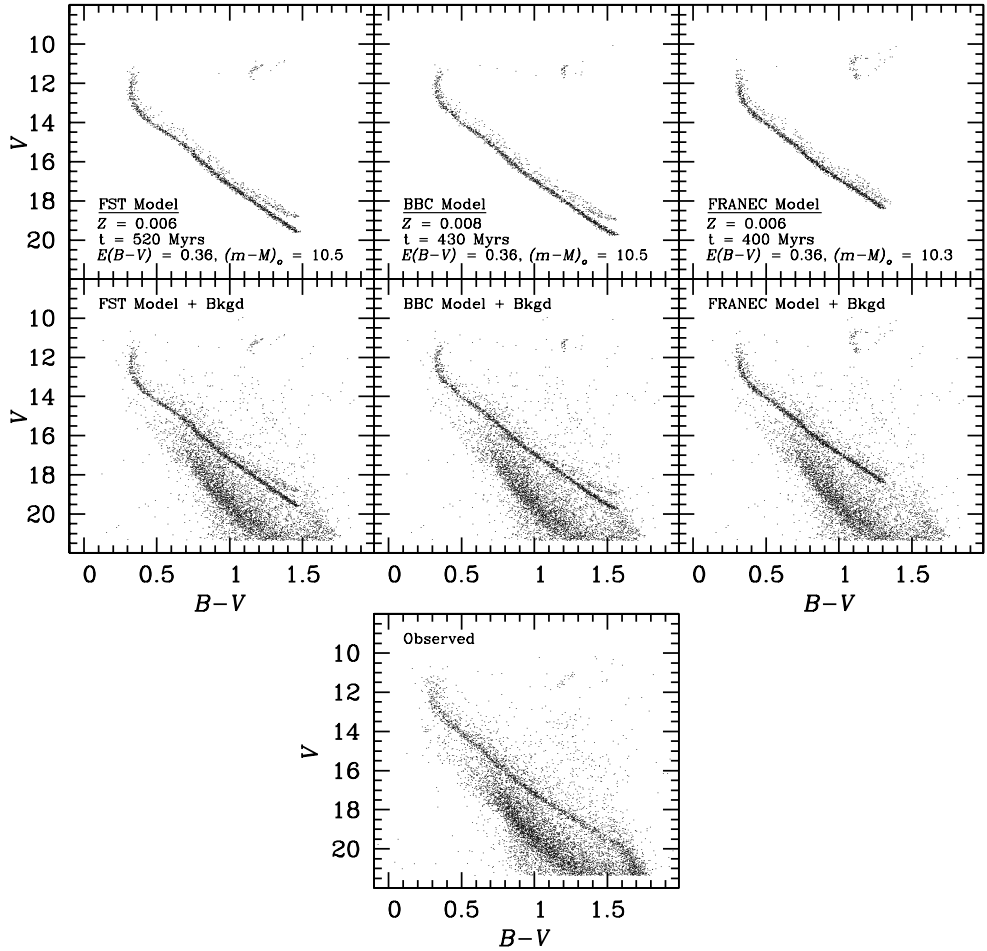,width=0.75\textwidth}
\caption{Synthetic CMDs with subsolar metallicities 
and a 20\% binary fraction (random mass ratio), are shown for NGC 2099.  
The BBC94 tracks are found to provide the best agreement with the observations 
(bottom panel). The data has been truncated at the faintest magnitudes (see Figure 1 
for full CMD).  See \S \ref{ngc2099} for more information.}
\label{N2099cmd2}
\end{figure*}


The cluster field contains 12194 {\it bona fide} stars, while the blank field 
extrapolated to the cluster area contains 10576 stars. Hence, the
synthetic CMDs have been computed with 1618 objects.

In JSKIII, we assumed the cluster metallicity of NGC 2099 to be Solar, based 
on the isochrone fitting results of \cite{mermilliod} as well as our own best 
guess from the fit.  A more recent study, \cite{nilakshi}, prefer a subsolar 
Z = 0.008 metallicity, again based on isochrone fitting.  Therefore, one goal 
is to constrain the metallicity of the cluster based on synthetic CMD 
comparisons.  So, for this cluster we use the the FST tracks of \cite{ventura} 
with $\zeta$ = 0.03, $Z$ = 0.02 and $Z$ = 0.006, the BBC94 tracks of the Padova group 
with $Z$ = 0.02 \citep{bressan} and $Z$ = 0.008 \citep{fagotto}, and the FRANEC tracks 
\citep{dominguez} with $Z$ = 0.02, $Z$ = 0.01 and $Z$ = 0.006.

For the FST, the synthetic sequences using Solar metallicity models look 
impressively consistent with the observed CMD in all the minimal details. However, the 
turn-off and clump shape are not very good: the sequence after the turn-off is too long and 
curved and the clump is a bit short and faint. It is also very difficult to single 
out the best age, as anything between 520 and 700 Myrs doesn't significantly change 
the important CMD features (clump morphology and luminosity, shape of turn-off, etc...), 
all looking reasonable.  The clump becomes too faint below 500 Myrs, indicating that 
this is a lower age limit and too blue beyond 800 Myrs indicating an upper limit. The 
best fit resulting parameters for these tracks are therefore age = 620 $\pm$ 60 Myrs, 
$E(B {\rm-}V)$ = 0.21 $\pm$ 0.02, and $(m {\rm-}M)_\circ$ = 10.70 $\pm$ 0.20.  The BBC94 
$Z$ = 0.02 tracks excellently reproduce all of the main sequence and clump 
features, but have long post turn-off sequences, as we saw in the FST models.   For Solar 
metallicity, these tracks are in better agreement with the data than the FST models. The 
age is also better constrained, thanks to a higher sensitivity of the various features: age 
= 590 $\pm$ 10 Myrs, $E(B {\rm-}V)$ = 0.18 $\pm$ 0.01, and $(m {\rm-}M)_\circ$ = 10.50 $\pm$ 
0.10.  The FRANEC $Z$ = 0.02 tracks provide the best agreement with the observed
upper main sequence and turn-off features. However, the resulting clump is too vertical 
and overpopulated.  Part of the latter problem is due to the fact that their main 
sequence is not as deep as the others, because the minimum mass available to us is 0.7 
M$_\odot$, instead of 0.6 M$_\odot$ as for the FST and Padova models.  The code therefore 
can not distribute the required 1618 faint stars in these bins and, inevitably, puts a 
larger number of brighter stars according to a Salpeter IMF.  The resulting 
best fit parameters in the FRANEC models are age = 590 $\pm$ 10 Myrs, $E(B {\rm-}V)$ 
= 0.16 $\pm$ 0.02, and $(m {\rm-}M)_\circ$ = 10.31 $\pm$ 0.10.  


\begin{figure*}
\centering
\epsfig{file=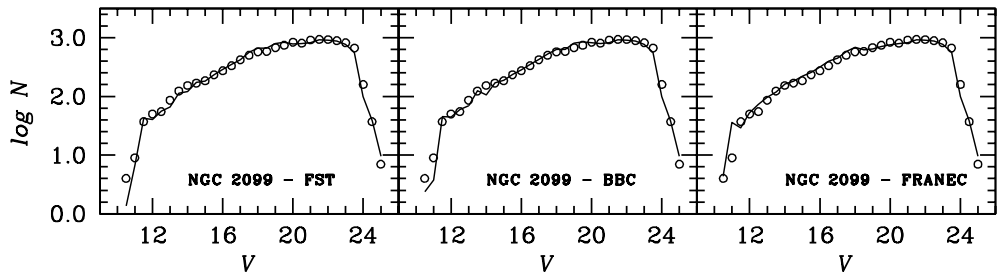,width=0.75\textwidth}
\caption{The empirical NGC 2099 luminosity function (open circles) 
is compared with the predicted luminosity function from the numerical simulations 
(solid line). See \S \ref{ngc2099} for a discussion of these results.}
\label{N2099lf}
\end{figure*}


Figure 4 presents the lower metallicity synthetic CMDs for NGC 2099.  For the $Z$ = 0.006 
FST models (left), the length of the sequence just after the turn-off shortens and becomes 
more vertical, thus better reproducing the observed morphology. The clump morphology 
also improves, thanks to a slightly larger extension. However, the agreement in the 
main-sequence morphology is no longer as excellent as the higher metallicity case, 
but still quite good.  It may well be that intermediate metallicity tracks would 
work even better.  Therefore, overall the synthetic CMDs from the FST tracks prefer 
the lower metallicity, with parameters: age = 520 $\pm$ 40 Myrs, $E(B {\rm-}V)$ = 
0.36 $\pm$ 0.01, $(m {\rm-}M)_\circ$ = 10.40 $\pm$ 0.10.  The BBC94 lower metallicity 
tracks (middle), $Z$ = 0.008, are the best models for this cluster. They reproduce 
all of the main-sequence features and also have a better shaped turn-off than with 
the Solar metallicity models.  A minor inconsistency is seen in the location of 
the clump, which is slightly too vertical, but this is a minor detail.  
The age is less well defined, with final best parameters age = 430 $\pm$ 30 Myrs, 
$E(B {\rm-}V)$ = 0.36 $\pm$ 0.03, $(m {\rm-}M)_\circ$ = 10.50.  The lower metallicity 
FRANEC tracks (right) can be ruled out as both the $Z$ = 0.01 and $Z$ = 0.006 models 
show extremely extended clumps (real blue loops) which are inconsistent with the 
observed clump for any reasonable age.  Also, the lifetimes in that phase favor 
the stars to reside on the blue side of the clump (vertical), adding further 
inconsistency to the colour and shape.

We have also perfomed several tests on the fraction of binary stars in NGC 2099. 
At first glance, the empirical CMD doesn't show any evidence of binaries, but 
we have found a posteriori that, without including them in the synthetic CMDs, 
both the turn-off morphology and the colour distribution on the red side of the main 
sequence are not well reproduced.  Our best guess is that binaries are around 
15-20\% of the cluster stars and all of the results that we have discussed above assume 
this fraction (20\%) with a random mass ratio.

Figure 5 shows the empirical luminosity function of the 12194 cluster field stars (empty 
circles) compared to the luminosity functions resulting from the sum of the 1618 synthetic 
stars of Figure 4 (subsolar metallicity) with the 10576 stars of the extrapolated 
blank field.  The predicted luminosity functions are consistent with the data, but 
the agreement at the brightest magnitudes is not exceptional. The very good match at the
faintest magnitudes shows that the extrapolation of the number of stars
observed in the blank field to the larger area covered by the cluster is
correct. 

The general fits of the data to synthetic CMDs favor the lower metallicity 
models for NGC 2099, thus requiring a higher reddening.  These are found to 
better reproduce the shape of the upper main sequence and turn-off.  Notice 
that age and modulus derived here do not coincide with those derived 
JSKIII with the same stellar tracks. The difference in age is probably due to 
the fact that with the isochrone fitting method one simply fits a curve to the 
points, whereas with the synthetic CMDs one has to reproduce the points 
themselves, including morphology and quantity.  The difference in modulus is, 
presumably, simply the consequence of the age difference.

Assuming a single slope Salpeter IMF, the mass of gas that formed the stars of
the surveyed area of NGC 2099  between 0.6 and 100 $M_{\odot}$ is 
$\simeq 2.6 \times 10^3 M_{\odot}$ with the BBC94 models, $2.5 \times 
10^3 M_{\odot}$ and $3.1 \times 10^3 M_{\odot}$ with the FST and the FRANEC
models, respectively.

\subsection{NGC 2168} \label{ngc2168}


\begin{figure*}
\centering
\epsfig{file=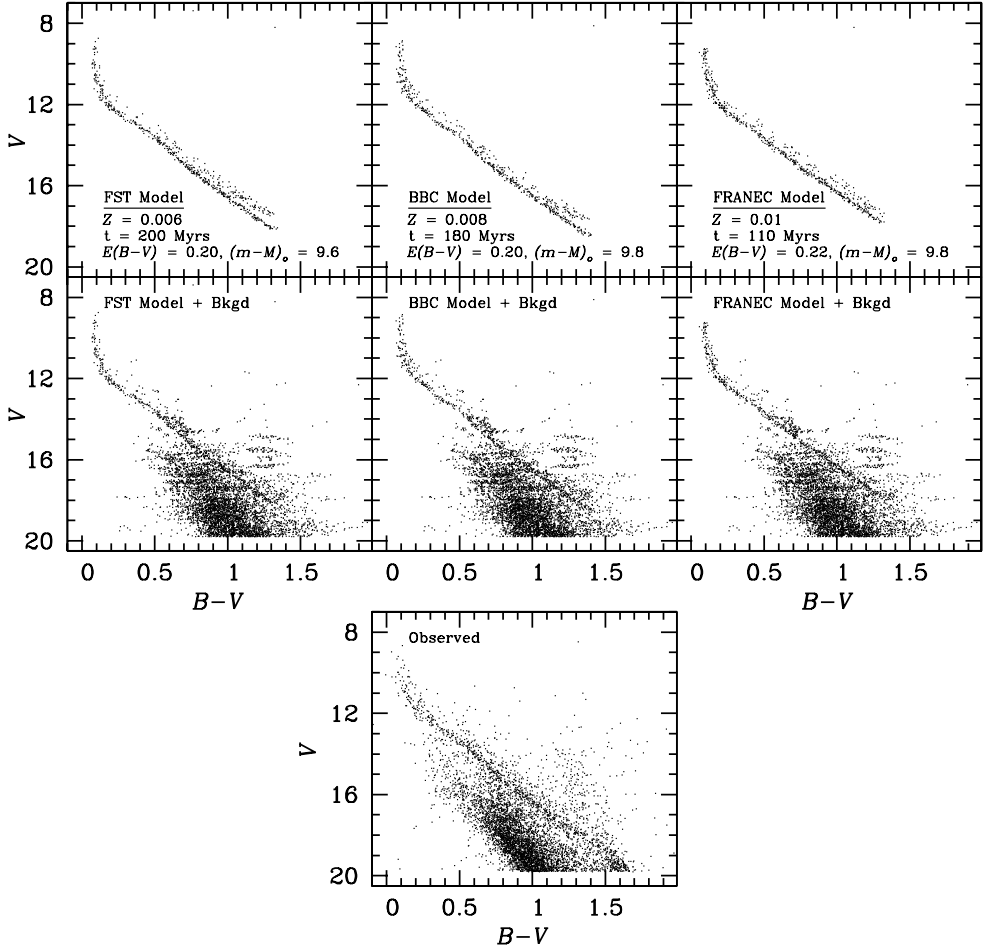,width=0.75\textwidth}
\caption{Synthetic CMDs with subsolar metallicities 
and a 30\% binary fraction (random mass ratio), are shown for NGC 2168.  
The BBC94 tracks are found to provide the best agreement with the observations 
(bottom panel).  The data has been truncated at the faintest magnitudes (see 
Figure 1 for full CMD).  See \S \ref{ngc2168} for more information.}
\label{N2168cmd}
\end{figure*}


The cluster field contains 9298 {\it bona fide} stars. Our previous work on 
NGC 2168 (JSKIV) suffered due to the large size of the cluster 
(R $>$ 20$'$).  A very small blank field was constructed from the outer edges of 
the CFH12K CCD mosaic.  In fact, some cluster members surely resided within the 
blank field, and therefore the cluster luminosity function could have been 
underestimated.  Using synthetic CMDs, we find that we can only reproduce the 
observed stellar distribution in the CMD and the luminosity function if we use 500-600 
stars down to $V$ = 19.  Using the blank field from the outer edges of the CCD 
subtracts off too many stars and forces a poor agreement between the theory and the 
observations.  This confirms our initial suspicion that the cluster is larger 
than our aerial coverage. The synthetic CMDs shown here contain 550 stars
brighter than $V$ = 19.

For the synthetic CMD comparisons, we again use both Solar and subsolar 
metallicity models from each of FST, BBC94, and FRANEC.  We immediately find 
that the subsolar metallicity models provide a much nicer agreement to the 
observations than Solar metallicity models.  The Solar metallicity models not 
only predict an insufficient number of main-sequence stars just below the 
turn-off, they also contain very vertical upper main sequences. Besides, for
ages as old as $\sim$180 Myrs the turn-offs are slightly too faint, while
for younger ages, $<$120 Myrs, the turn-off has the proper luminosity but
is severely underpopulated when compared to the observations.  
Flattening the IMF did not help the overall fit.

Figure 6 shows the best synthetic CMDs for NGC 2168.  For the FST tracks (left), 
the models with maximum overshooting ($\zeta$ = 0.03) and low metallicity 
($Z$ = 0.006) reproduce the data very nicely.  These require a best fit distance 
modulus slightly smaller ($(m {\rm-}M)_\circ$ = 9.6) and a reddening slightly 
larger ($E(B {\rm-}V)$ = 0.22) than found in JSKIV.  The differences can be most 
likely attributed to the lower metallicity (JSKIV uses $Z$ = 0.012) that we chose 
in these comparisons.  The best age for this model is 200 Myrs, slightly older 
than what JSKIV found (180 Myrs).  The absence of a population of post 
main-sequence stars coupled with the insensitivity of the turn-off luminosity 
and morphology with age in these models allows for a nice agreement between the 
theory and data with ages down to 120 Myrs.
            
The agreement is found to be not as good if we use the same metallicity FST models 
with slightly lower overshooting, $\zeta$ = 0.02.  At an age of 
180 Myrs the turn-off is too faint, however, with younger ages (150 Myrs) we 
overpopulate the post main-sequence stars (five are seen).  At 120 Myrs, the 
turn-off is too bright, hence this age is too young.


\begin{figure*}
\centering
\epsfig{file=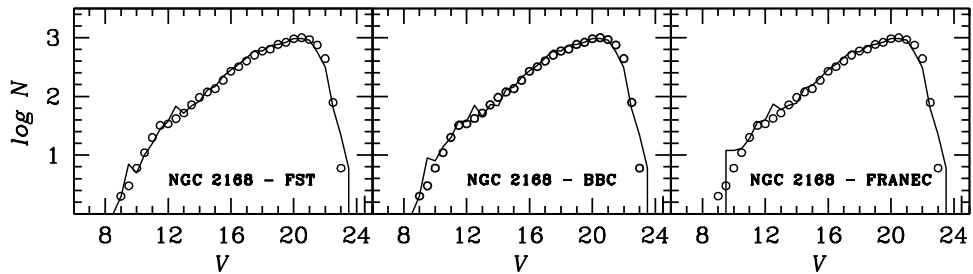,width=0.75\textwidth}
\caption{The empirical NGC 2168 luminosity function (open circles) 
is compared with the predicted luminosity function from the numerical simulations 
(solid line). See \S \ref{ngc2168} for a discussion of these results.}
\label{N2168lf}
\end{figure*}


The BBC94 models for NGC 2168 are shown in the middle panel of Figure 6.  The 
$Z$ = 0.08 tracks are found to excellently reproduce both the CMD morphology and 
the number of stars in the different phases. The main-sequence curves are best 
fitted using $E(B {\rm-}V)$ = 0.20 and $(m {\rm-}M)_\circ$ = 9.8. The best age 
estimate is found to be 180 Myrs.  An age of 150 Myrs doesn't provide any post 
main-sequence stars whereas an age of 200 Myrs is too old as it predicts 
too many clump stars.  The FRANEC models (right) predict too many post 
main-sequence stars for both adopted metallicities, $Z$ = 0.01 and $Z$ = 0.02.  
As summarised in JSKIV, many studies of NGC 2168 exist in the literature with 
none finding more than a few clump stars.  The FRANEC models would require a 
very young age (110 Myrs) to fit the post main-sequence phases and this would 
in turn force a poor agreement in the turn-off region.  The best fit model is, 
arguably, the one with age = 110 Myrs, $E(B {\rm-}V)$ = 0.22, and 
$(m {\rm-}M)_\circ$ = 9.8.

Our best estimate for the NGC 2168 binary fraction is found to be $\sim$30\%.  
This fraction puts a consistent number of stars, and sparse regions, on the 
red side of the cluster main-sequence in the synthetic CMDs.  All CMDs shown in 
Figure 6 contain this binary fraction.  

Comparing the resulting best synthetic CMDs with the observed one, 
it is clear that the predicted main sequence is too tight and slightly 
underpopulated at faint magnitudes ($V >$ 15).  As addressed earlier in 
\S \ref{ngc6819}, the tightness is a question of photometric errors, possible 
differential reddening, or metallicity spreads within the cluster.  
The underpopulation in this case is most likely related to the uncertain method 
we have used to infer the number of cluster members.  Many tests that we have 
done indicate that we cannot attribute more stars to the cluster without 
getting too many bright stars, both on the main sequence and in subsequent 
evolutionary phases.  We also note here that $V$ = 19 is the limiting magnitude 
of the numerical simulations for this cluster.  This is related to the 
availability of photometric conversion tables \citep{bessell} for the adopted 
metallicities.  With that said, it is remarkable how well the overall shape 
of the upper main sequence is reproduced.  The 'kinks' and slope changes all 
fall in the correct place as dictated by the observations.  This allows us to 
accurately determine the best fit parameters for NGC 2168.  Interestingly, 
the best fit synthetic CMD for this cluster, the BBC94 model, provides an 
identical reddening, distance, and age to what we found using the FST isochrone 
in JSKIV.  The FST synthetic CMD derived parameters are also in excellent 
agreement with our observations considering that the metallicities are slightly 
different.

Figure 7 shows the empirical luminosity function of the 9298 cluster field stars (empty circles) 
compared to the luminosity functions resulting from the sum of the synthetic stars of 
Figure 6 with the 8748 stars resulting by randomly multiplying the 1150 objects
of the supposed blank field for an area normalisation factor. Clearly, a major 
side effect of the lack of an appropriate external field is that
the comparison of the luminosity functions becomes only a matter of self-consistency and
cannot be used for any consideration on mass segregation/evaporation.
As mentioned above, the luminosity functions agree with the data only with 
500-600 cluster members. One may think to improve the situation by steepening 
significantly the IMF (thus allowing for more stars in the fainter main sequence, without 
adding bright ones), but we consider it highly dangerous to modeify the IMF 
without a good field subtraction.  We stress that observations of a deep $BV$ field 
offset from the centre of NGC 2168 by at least one degree would be very useful.


\begin{figure*}
\centering
\epsfig{file=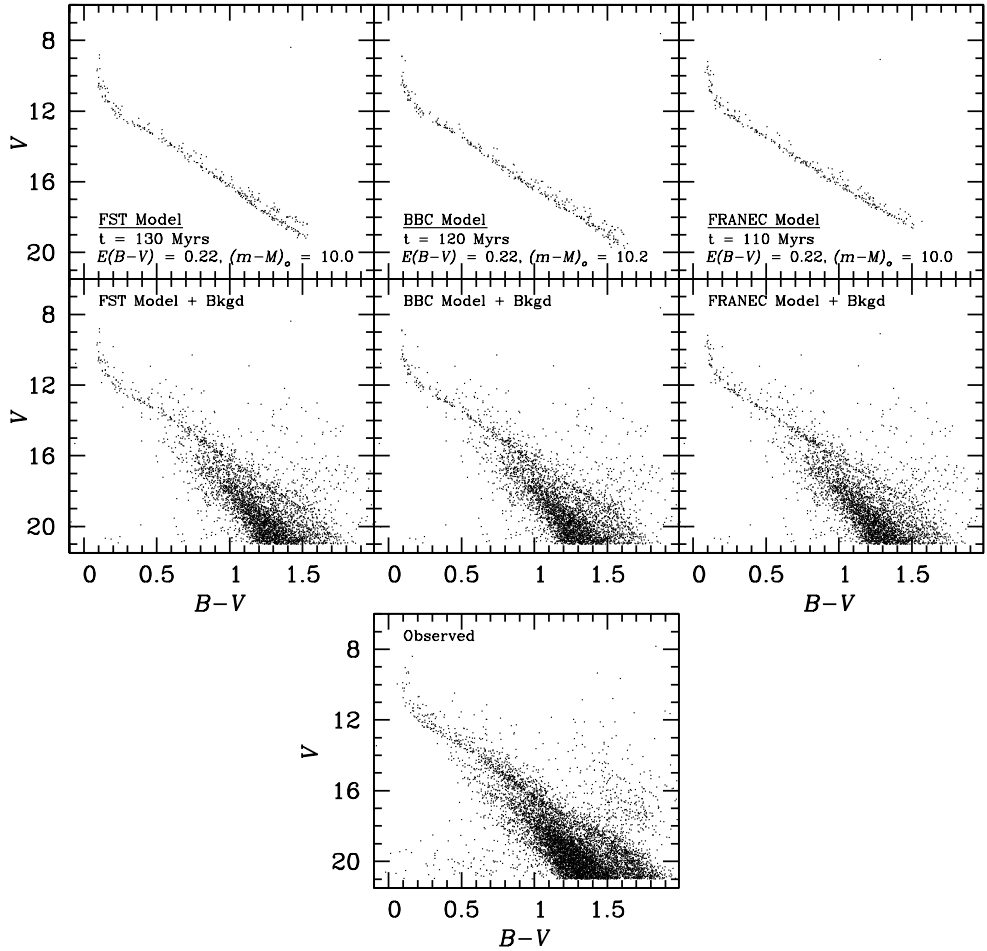,width=0.75\textwidth}
\caption{Synthetic CMDs with $Z$ = 0.02 and a 30\% 
binary fraction (random mass ratio), are shown for NGC 2323.  
The BBC94 tracks are found to provide the best agreement with the observations 
(bottom panel).  The data has been truncated at the faintest magnitudes (see 
Figure 1 for full CMD).  See \S \ref{ngc2323} for more information.}
\label{N2323cmd}
\end{figure*}


Assuming a single slope Salpeter IMF, the mass of gas that formed the stars of
the surveyed area of NGC 2168 between 0.6 and 100 $M_{\odot}$ is 
$\simeq 8.5 \times 10^2 M_{\odot}$ with the BBC94 models, $7.8 \times 
10^2 M_{\odot}$ and $9.6 \times 10^2 M_{\odot}$ with the FST and the FRANEC
models, respectively.

\subsection{NGC 2323} \label{ngc2323}

The cluster field contains 11115 {\it bona fide} stars. Taking into account
that 1) the blank field contains 5142 stars but covers an area 1.8 times 
smaller than the cluster field, 2) the mass of the available stellar models
don't allow synthetic stars fainter that $V \sim 18$, the
synthetic CMDs have been initially computed with 565 objects brighter than $V$ = 18.
A posteriori, we have however found that in this way we predict luminosity functions 
clearly overestimating the bright portion (where the cluster dominates) and 
underestimating the faint portion (where field contamination dominates). We
empirically find that good luminosity functions are obtained only if the cluster contains
285 members brighter than $V = 18$, suggesting that the previously used blank field 
underestimated the actual contamination. Again observations of an external 
field of appropriate size would be crucial for a safer derivation of some of 
the cluster properties.

For the youngest cluster in this work, NGC 2323, the 
metallicity is known to be Solar and in fact, we find that the Solar 
metallicity tracks provide the best agreement to the data in this work.  The 
most difficult parameter to fit in NGC 2323 is the cluster age.  Although 
ages can be constrained from the turn-off morphology, the luminosity of the 
red giant clump is an alternative tool.  NGC 2323 only shows two bright, 
red objects which are potential post main-sequence candidates.  However, 
since these 
two objects do not have any membership probability from the literature, we can 
not trust them as cluster members.  Furthermore, the turn-off itself is not 
heavily populated preventing an accurate determination of its luminosity.  
As we'll see below, we find that models with a range of young ages are 
equally consistent with the observed distribution.

Figure 8 presents our best synthetic CMDs for NGC 2323.  All of the shown cases 
are for Solar metallicity and assume a 30\% binary fraction with random mass 
ratio.

The FST models (left) are found to perfectly reproduce the majority of the 
observed features.  The agreement is also found to be independent of the 
adopted overshooting ($\zeta$ = 0.02 or $\zeta$ = 0.03).  Although the age 
is slightly higher (by $\sim$20 Myrs) in the higher overshooting case, the reddening, 
distance modulus and goodness of fit are the same.  Our best parameters are 
therefore age = 130 Myrs, $E(B {\rm-}V)$ = 0.22, and $(m {\rm-}M)_\circ$ = 
10.0 for $\zeta$ = 0.03.  However, ages between 120 and 180 Myrs are equally 
likely.  We have chosen 130 Myrs as this age predicts a post main-sequence star 
where a real star (member or not) is indeed located on the CMD.   Therefore, the 
results from this work for the FST models are identical to the parameters found 
in JSKIV.  The BBC94 stellar tracks also perfectly reproduce features on the CMD of 
this cluster.  For $E(B {\rm-}V)$ = 0.22, the best fit distance modulus is 
found to be $(m {\rm-}M)_\circ$ =  10.2.  In practice, all ages between 80 and 
150 Myrs fit the data equally well.  We've again chosen 120 Myrs as the best 
age as it predicts a post main-sequence star in the correct location as dictated 
by the potential red giant in the data.  The FRANEC models (right) predict a larger 
number of post main-sequence stars than the FST and BBC models.  This is 
therefore inconsistent with the observational data.  For Solar metallicity, the best 
compromise between the number of post main-sequence stars, the turn-off luminosity, 
and the shape of the upper main sequence is found for an age of 110 Myrs.  The turn-off 
is still too faint, however, younger ages produce very vertical and blue upper main 
sequences.  The best fit reddening is $E(B {\rm-}V)$ = 0.22 and the distance is 
$(m {\rm-}M)_\circ$ =  10.0.  Since the results from these latter comparisons were 
not very satisfactory, we also tried a lower metallicity, $Z$ = 0.01.  The results are 
very similar, with an overproduction of post main-sequence stars and a worse shape 
to the upper main sequence.


\begin{figure*}
\centering
\epsfig{file=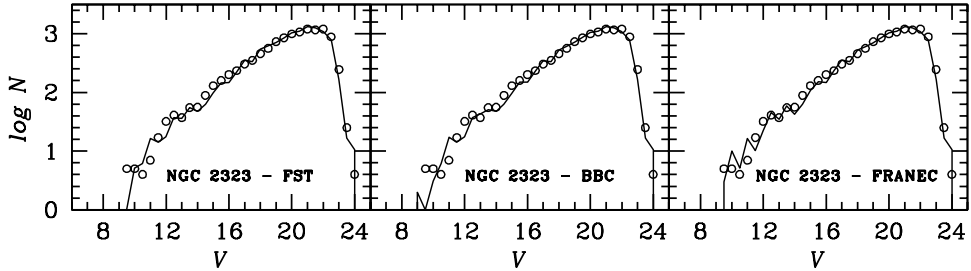,width=0.75\textwidth}
\caption{The empirical NGC 2323 luminosity function (open circles) 
is compared with the predicted luminosity function from the numerical simulations 
(solid line). See \S \ref{ngc2323} for a discussion of these results.}
\label{N2323lf}
\end{figure*}


Figure 9 shows the empirical luminosity function of the 11115 cluster field stars (empty 
circles) compared to the luminosity functions resulting from the sum of the synthetic 
stars of Figure 8 with the 10576 stars of the extrapolated blank field. As already
discussed for NGC 2168, the lack of an appropriate estimate of the
back/foreground stars doesn't allow for a critical analysis of the luminosity function. 
We can see, however, from Figure 8, that within the uncertainties, the selected
models reproduce quite well the observed luminosity function.

Assuming a single slope Salpeter IMF, the mass of gas that formed the stars of
the surveyed area of NGC 2323 between 0.6 and 100 $M_{\odot}$ is 
$\simeq 4.8 \times 10^2 M_{\odot}$ with the BBC94 models, $4.3 \times 
10^2 M_{\odot}$ and $5.0 \times 10^2 M_{\odot}$ with the FST and the FRANEC
models, respectively.

\section{Discussion} \label{discussion}

Table 2 summarises the results described in the previous section and includes 
other information about the four clusters studied here.  We recall that 
stellar models 
that include overshooting from convective regions imply older ages than models 
without overshooting. This explains why our ages resulting from the FST tracks 
with $\zeta$=0.03 are systematically older than those obtained with the BBC94 
models and much older than those obtained 
with the FRANEC ones. It is interesting to notice that while the FST 
models with higher overshooting reproduce the data better than those with 
$\zeta$=0.02, the BBC94 models appear to be those in  better overall
agreement, in spite of having a formal overshooting somewhat lower than the
FST $\zeta$=0.03 models. This indicates that what counts in the predictions of
stellar evolution models is the combination of the many parameters and input
physics. We find that in general the BBC94 models better reproduce both the
morphological features and the numbers of stars in the various evolutionary
phases of the observed clusters.

\begin{table*}
\caption{Summary of Parameters and Results}
\begin{tabular}{llllll}
\hline
Parameter & Explanation & NGC 6819 & NGC 2099 & NGC 2168 & NGC 2323 \\
\hline

Position: & \\
$\alpha_{\rm J2000}$  & RA                          & $19^{\rm h}41^{\rm m}17.7^{\rm s}$ & $06^{\rm h}08^{\rm m}54.0^{\rm s}$ & $06^{\rm h}08^{\rm m}54.0^{\rm s}$ & $07^{\rm h}02^{\rm m}48.0^{\rm s}$    \\ 
$\delta _{\rm J2000}$ & declination                 & $+40^{\rm o}11'17''$               & $+24^{\rm o}20.0'$                 & $+24^{\rm o}20.0'$                 & $-08^{\rm o}22'36''$                  \\
$l_{\rm J2000}$       & Galactic longitude          & $73.98^{\rm o}$                    & $177.65^{\rm o}$                   & $186.58^{\rm o}$                   & $221.67^{\rm o}$                      \\   
$b_{\rm J2000}$       & Galactic latitude           & $8.47^{\rm o}$                     & $3.09^{\rm o}$                     & $2.18^{\rm o}$                     & $-1.24^{\rm o}$                       \\ 
\\	                     
Distances: & \\				                                                                                                              	                                       
$(m {\rm-}M)_{V}$ - Previous Study       & apparent distance modulus   & 12.30 $\pm$ 0.12        & 11.55 $\pm$ 0.13        & 10.42 $\pm$ 0.13        & 10.68 $\pm$ 0.14         \\
$(m {\rm-}M)_{V}$ - FST ($\zeta$ = 0.03) & apparent distance modulus   & 12.31                   & 11.52                   & 10.22                   & 10.68                    \\
$(m {\rm-}M)_{V}$ - BBC94                & apparent distance modulus   & 12.57                   & 11.62                   & 10.42                   & 10.88                    \\
$(m {\rm-}M)_{V}$ - FRANEC               & apparent distance modulus   & 12.47                   & 11.42                   & 10.48                   & 10.68                    \\
\\
$E(B {\rm-}V)$    - Previous Study       & reddening                   & 0.10                    & 0.21                    & 0.20                    & 0.22                     \\
$E(B {\rm-}V)$    - FST ($\zeta$ = 0.03) & reddening                   & 0.10                    & 0.36                    & 0.20                    & 0.22                     \\
$E(B {\rm-}V)$    - BBC94                & reddening                   & 0.12                    & 0.36                    & 0.20                    & 0.22                     \\
$E(B {\rm-}V)$    - FRANEC               & reddening                   & 0.15                    & 0.36                    & 0.22                    & 0.22                     \\
\\
$(m {\rm-}M)_\circ$ - Previous Study          & true distance modulus  & 11.99 $\pm$ 0.18        & 10.90 $\pm$ 0.16        & 9.80 $\pm$ 0.16         & 10.00 $\pm$ 0.17          \\
$(m {\rm-}M)_\circ$ - FST ($\zeta$ = 0.03)    & true distance modulus  & 12.0                    & 10.40                   & 9.60                    & 10.00                     \\
$(m {\rm-}M)_\circ$ - BBC94                   & true distance modulus  & 12.2                    & 10.50                   & 9.80                    & 10.20                     \\
$(m {\rm-}M)_\circ$ - FRANEC                  & true distance modulus  & 12.0                    & 10.30                   & 9.80                    & 10.00                     \\
\\
$d$ - Previous Study          & distance from Sun                   & 2500 $\pm \ ^{216}_{199}$ pc  & 1513 $\pm \ ^{146}_{133}$ pc  & 912 $\pm \ ^{70}_{65}$ pc    & 1000 $\pm \ ^{81}_{75}$ pc            \\
$d$ - FST ($\zeta$ = 0.03)    & distance from Sun                   & 2512 pc       & 1202 pc       & 832           & 1000       \\
$d$ - BBC94                   & distance from Sun                   & 2754 pc       & 1259 pc       & 912           & 1096       \\
$d$ - FRANEC                  & distance from Sun                   & 2512 pc       & 1148 pc       & 912           & 1000       \\
\\							                                                                                                              	                                       
Age: & \\						                                                                                                              	                                       
$t$ - Previous Study        & isochrone fit age   & 2.5 Gyrs                           & 520 Myrs                           & 180 Myrs                           & 130 Myrs   \\
$t$ - FST ($\zeta$ = 0.03)  & synthetic CMD age   & 2.3 Gyrs                           & 520 Myrs                           & 200 Myrs                           & 130 Myrs   \\
$t$ - BBC94                 & synthetic CMD age   & 2.0 Gyrs                           & 430 Myrs                           & 180 Myrs                           & 120 Myrs   \\
$t$ - FRANEC                & synthetic CMD age   & 1.6-1.8 Gyrs                       & 400 Myrs                           & 110 Myrs                           & 110 Myrs   \\
\\							                                                                                                              	                                       
Metallicity: & \\					                                                                                                              	                                       
$Z$           & heavy metal abundance$^{\rm 1}$       & 0.02                               & $<$0.02                              & 0.012                               & 0.020                                \\
\\							                                                                                                            	                                       
Binary Fraction & \\
Binaries              & binary percentage           & 20\%  & 20\% & 30\% & 30\% \\
\\
Size: & \\						                                                                                                              	                                       
$\Theta$  & angular diameter                        & 19$'$                              & 27$\farcm$8                       & $>$32$'$                           & 30$'$                                 \\
$D$       & linear diameter$^{2}$                   & 15.2 pc                            & 10.2 pc                           & $>$8.5 pc                          & 9.6 pc                                \\
$M$       & mass of cluster$^{3}$                   & 4000 $M_\odot$                     & 2500 $M_\odot$                    & 850 $M_\odot$                      & 480 $M_\odot$                         \\
\\							                                

\hline

\end{tabular}

1. These values were not spectroscopically determined and reflect those used in the original isochrone fits. \\
2. Computed using the distances from the BBC94 results. \\
3. M is the mass of gas that formed stars between 0.6 and 100 $M_\odot$ and is therefore only a lower limit to the total astrated mass. \\

\end{table*}

The excellent delineation of the various main-sequence curves and kinks allows 
us to put
fairly stringent limits on both the reddening and the distance modulus of each
cluster, independent of the age. Table 2 shows in fact that the variation of
the values of these two quantities resulting from different sets of stellar
tracks is quite small. This clearly helps to better define the age as well.

As discussed for the case of NGC 6819, the synthetic main sequences of all the
four clusters are tighter than the observed main sequences when the formal 
photometric
errors of the survey catalogues are assumed. To reproduce the observed width,
the synthetic CMDs require either photometric errors of at least 0.01 magnitude 
even in the brighter bins or some other factors such as an internal metallicity 
spread or differential reddening.

The synthetic CMD tests clearly indicate that NGC 2168 is larger than previously 
thought and that our outer CCD blank field is not adequate.  A similar problem, 
although to a lesser degree, may also exist in the NGC 2323 data set.  Future 
deep photometry of these clusters should include separate blank field 
observations. Indeed, in spite of the great depth and excellent accuracy of 
our photometry, which allows us to measure the faintest, least massive stars 
of each cluster, the lack of appropriate decontamination fields has prevented 
us from safely deriving their present mass functions and, 
consequently, their IMF, mass segregation, and evaporation status.

The synthetic CMDs have allowed us to infer the astrated mass of each cluster.
We find that although this does depend on the adopted set of stellar evolutionary
tracks, it only varies within  a factor 1.2 from the minimum to the maximum 
value obtained with the stellar models considered here. 
We have chosen to simply apply Salpeter's single slope IMF for this 
exercise. We note, however, that increasing evidence suggests that the slope 
is most likely different (steeper for more massive stars, flatter for less 
massive ones, e.g., Kroupa 2002) beyond the range of masses, 
$2 - 10 M_{\odot}$, originally analysed by Salpeter (1955).  
Although this simplistic assumption introduces a systematic error in our 
estimates, it is not a large concern as we are not providing absolute 
values of star formation rates but rather just comparing relative rates for the 
four clusters.  

Given the angular size of the observed field of view for each 
cluster (see Table 2), the actual area surveyed obviously depends 
on the adopted distance to the cluster.  We have assumed both the instrinsic 
distance modulus and the astrated mass resulting from the best synthetic 
CMD comparisons, the BBC94 models. This corresponds to a distance from the 
Sun of 2754 pc for NGC 
6819, 1259 pc for NGC 2099, 912 pc for NGC 2168, and 1096 pc for NGC 2323. The 
surveyed areas are then 182, 81, 56 and 72 $pc^2$, respectively. The 
corresponding astrated mass per unit area is then 23.1, 32.1, 15.2 and 
6.7 $M_{\odot}pc^{-2}$, equivalent to star formation rates of 
$4.6 \times 10^{-6}, 6.4 \times 10^{-6}, 3.0 \times 10^{-6}$, and 
$1.3 \times 10^{-6} M_{\odot}yr^{-1}pc^{-2}$, respectively. 
These values refer only to stars with masses included in the adopted sets of
evolutionary tracks, namely from 0.6 to 100 $M_{\odot}$. To derive
total astrated masses and star formation rates, one should extrapolate
down to 0.1 $M_{\odot}$, following the preferred IMF. In practice,
extrapolating Salpeter's IMF down to 0.1 $M_{\odot}$, we must multiply by a
factor 2.05 the astrated masses and the rates. If, at the other extreme, 
we prefer to assume the positive slope +0.44 below 0.6 $M_{\odot}$ suggested 
by Gould, Bahcall \& Flynn (1997), then the above values must be multiplied 
only by 1.27.

There is a significant difference (more than a factor of four) in the
star formation rates of our four clusters, the most active having been NGC
2099 and the least active NGC 2323, with no apparent relation with Galactic
location. These rates are 2--3 orders of magnitude higher than the average 
star formation rate of field stars in the Solar neighbourhood, 
($2 - 10) \times 10^{-9} M_{\odot}yr^{-1}pc^{-2}$ (e.g. Timmes, Woosley 
\& Weaver 1995), and are comparable to those of the most active dwarf irregular
and blue compact galaxies (e.g. Tosi 2003 and references therein).
The survival factor of the stars formed in
any cluster clearly depends both on age and on evaporation and NGC 6819
turns out to be the only one of the four clusters having suffered a reduction
from the number of formed stars to that of the stars still alive and present in the
studied area.

\section{Conclusions} \label{conclusions}

We re-derive key parameters for the four very rich open star clusters, 
NGC 6819, NGC 2099, NGC 2168, and NGC 2323.  The parameters are measured by 
comparing high quality, deep empirical colour-magnitude diagrams with 
MonteCarlo simulations.  
The combination of comparing the morphology and number of stars in various 
evolutionary phases and cluster luminosity functions allows us to provide 
tight constraints on the reddening, distance, age and binary fraction for each 
cluster.  In cases where the cluster metallicity is not certain, simulations 
with different abundances are also compared.  In all cases the data are better 
reproduced when a fraction of unresolved binary systems between 20 and 30\% is 
assumed. A summary of the results is given in Table 2.

The synthetic CMDs and LFs are generally found to be in excellent agreement 
with the observational data.  This circumstance, combined with the fact that
different sets of stellar evolution tracks provide different values for the 
cluster parameters, confirms how important it is to use more than one set of
models to estimate the theoretical uncertainties. It also shows that a
homogeneous approach is crucial to derive reliable overall cluster properties,
such as age-metallicity relations. In fact, cluster dating based on different
stellar models may lead not only to different absolute ages, but also to
different age ranking, with significant drawbacks on the interpretation of the
cluster properties in terms of Galactic evolution.

Some discrepancies, such as the thickness of the main sequences {\it vis \`a
vis} the size of the photometric errors, and the possibility that cluster 
stars may be contaminating our blank field (particularly in the case of 
NGC 2168) are identified and discussed. 
 
For each cluster, we also measure the astration mass (i.e., the total mass that
went in all of the stars formed in the cluster) according to a single slope
Salpeter IMF.  From this, we calculate the star formation rate between 
0.1--100 $M_{\odot}$ to be 
$9.4 \times 10^{-6}, 1.3 \times 10^{-5}, 6.2 \times 10^{-6}$, and 
$2.7 \times 10^{-6} M_{\odot}yr^{-1}pc^{-2}$
for NGC 6819, NGC 2099, NGC 2168, and NGC 2323 respectively. These
rates would be a factor 1.6 lower if the IMF below 0.6 $M_{\odot}$ has a 
slope of +0.44, as inferred by Gould et al. (1997) from HST data, rather than 
Salpeter.  The true value may in fact lie somewhere in between these two IMFs, as 
recently discussed e.g., by Chabrier (2003). In principle our data are deep enough 
to allow for a direct derivation of the cluster IMF; however, the four clusters 
are heavily contaminated by fore/background stars and the lack of appropriate
decontamination fields has prevented us from a safe analysis of the star counts
at the fainter magnitudes. The need of appropriate photometry in
nearby fields is emphasised, also for the purpose of adequate studies of the
clusters mass segregation/evaporation.

\section*{Acknowledgments}

We thank Paolo Ventura and Franca D'Antona for having provided their stellar
models prior to publication.  The bulk of the synthetic CMD code was originally 
written by Laura Greggio.  J. S. K. received financial support during this work 
through an NSERC PGS-B graduate student research grant.  This work has been 
partially funded through the Italian MIUR-Cofin-2003029437.

\bsp

\label{lastpage}

\end{document}